\documentclass[sort&compress,10pt,twocolumn,final,3p]{elsarticle}

\usepackage[utf8]{inputenc}
\usepackage{hyperref}
\usepackage{graphicx}
\usepackage{subfigure}
\usepackage{array}
\usepackage{multirow}
\usepackage{amsmath,amssymb}
\usepackage{bm}	

\usepackage{miller}

\newcommand{\etal}{\textit{et al}.}
\newcommand{\ie}{\textit{i.e.}}

\newcommand{\lammps}{\textsc{Lammps}}
\newcommand{\ovito}{\textsc{Ovito}}
\newcommand{\insitu}{\textit{in situ}}
\newcommand{\Insitu}{\textit{In situ}}
\newcommand{\abinitio}{\textit{ab initio}}
\newcommand{\Abinitio}{\textit{Ab initio}}

\journal{Acta Materialia}

\begin{document}

\begin{frontmatter}

\title{Secondary slip of screw dislocations in zirconium}

\author[SRMP]{Émile Maras}
\author[SRMP]{Emmanuel Clouet\corref{CA}}
\cortext[CA]{Corresponding author}
\ead{emmanuel.clouet@cea.fr}
\address[SRMP]{Université Paris-Saclay, CEA, Service de Recherches de Métallurgie Physique, 91191, Gif-sur-Yvette, France}

\begin{abstract}
	Plasticity in hexagonal close-packed zirconium is controlled by screw dislocations which easily glide in the prismatic planes where they are dissociated.  At high enough temperatures, these dislocations can deviate out of the prism planes to also glide in the first order pyramidal and basal planes.  To get a better understanding of these secondary slip systems, we have performed molecular dynamics (MD) simulations of a screw dislocation gliding in a basal plane.  The gliding dislocation remains dissociated in the prism plane where it performs a random motion and occasionally cross-slips out of its habit plane by the nucleation and propagation of a kink-pair.  Deviation planes are always pyramidal, with an equal probability to cross-slip in the two pyramidal planes on both sides of the basal plane, thus leading to basal slip on average.  Basal slip appears therefore as a combination of prismatic and pyramidal slip in the high stress regime explored in MD simulations.  This is confirmed by nudged elastic band (NEB) calculations.  But NEB calculations also reveal a change of glide mechanism for a decreasing applied stress.  At low stress, kinks do not lie anymore in the pyramidal planes.  They are now spread in the basal planes, thus fully compatible with a motion of the screw dislocation confined to the basal plane as seen in experiments.  Basal slip, which is in competition with pyramidal slip, appears therefore favoured at low stress in pure zirconium. 
\end{abstract}


\end{frontmatter}

\section{Introduction}

Plasticity in hexagonal close-packed (hcp) zirconium mainly processes by glide of dislocations 
with $\hkl<a>=1/3\,\hkl<1-210>$ Burgers vector in the prismatic \hkl{10-10} planes. 
This principal slip system observed in hcp Zr is well rationalised by \abinitio{} calculations
which show that \hkl<a> screw dislocations are dissociated in the prismatic planes
where they can easily glide without experiencing any relevant lattice friction 
\cite{Clouet2012,Clouet2015}.
But at high enough temperatures, these dislocations can deviate out of the prism planes
to also glide in the basal \hkl(0001) planes, and more rarely in the pyramidal \hkl{10-11} planes.

Numerous transmission electron microscopy observations 
\cite{Bailey1962,Akhtar1971,Akhtar1973,Long2015a,Caillard2018}
have provided evidences of basal slip at various temperatures.
Post-mortem TEM \cite{Bailey1962,Long2015a} 
show the presence of \hkl<a> dislocations lying in the basal planes at room temperature 
after sufficient plastic strain, 
and Akhtar and Techtsoonian \cite{Akhtar1971}
observed that basal slip is already active at 78\,K when the strain is high enough.
At much higher temperatures, above 850\,K, Akhtar \cite{Akhtar1973} showed that 
basal slip was the only  activated secondary slip system in addition to easy prismatic slip,
with the relative ease of basal slip increasing with temperature.
Slip traces analysis also confirm the activation of basal slip at room temperature 
\cite{Dickson1971,Francillette1997,Francillette1998,Gong2015} 
and above up to 623\,K \cite{Wang2019}.
Finally, basal slip could be observed \insitu{} by TEM in pure Zr at and above room temperature
\cite{Caillard2015b,Caillard2018}.
In agreement with these observations, 
\insitu{} compression experiments during neutron diffraction experiments \cite{Long2013} 
on polycrystals with various textures
suggest basal slip activity operating at a higher applied stress than prismatic slip 
at room temperature, but at a similar applied stress above 373\,K.

\Insitu{} TEM observations \cite{Caillard2018} show that basal slip is controlled by the viscous glide 
of straight screw dislocations, an indication of a Peierls mechanism 
with dislocation moving by the nucleation and propagation of kink pairs.
This agrees with \abinitio{} calculations which show that a dissociation 
of the screw dislocation in the basal plane is unstable \cite{Clouet2012} 
and that screw dislocations remain dissociated in the prismatic plane 
while gliding in the basal plane, thus leading to a high energy barrier 
opposing basal glide \cite{Chaari2014}.

Although pyramidal glide of $\hkl<a>$ dislocations is another possible 
secondary slip system in hcp Zr, less experimental observations reported activity 
of this slip system.
Tenckhoff \cite{Tenckhoff2005} mentioned that pyramidal slip occurs in regions of high stress concentration,
such as grain boundaries,
and Caillard \etal{} \cite{Caillard2018} underlined that pyramidal slip necessitates
especially favourable stress conditions compared to basal slip.
Direct evidences of pyramidal slip have been indeed mainly detected in zirconium alloys 
and at high enough temperature.
Post-mortem TEM observations in a M5 alloy \cite{Rautenberg2012} reveal glide of \hkl<a> dislocations
in pyramidal \hkl{10-11} planes activated through cross-slip from prismatic principal slip systems
after creep at 673\,K.
But \insitu{} TEM straining experiments performed in the same alloy between 523 and 723\,K
indicate that cross-slip proceeds mainly in the basal plane \cite{Caillard2015b,Caillard2018}.
On the other hand, Drouet \etal{} \cite{Drouet2016} observed during \insitu{} TEM straining experiments 
glide of \hkl<a> dislocations in pyramidal planes in recrystallised Zircaloy-4
between 723 and 773\,K and reported that basal slip was not activated.
Pyramidal slip was also reported by Gaumé \etal{} \cite{Gaume2018} still in Zircaloy-4, 
both at room temperature and between 623 and 723\,K either in unirradiated or irradiated samples.

Basal and pyramidal slip appear therefore as two possible secondary slip systems
for \hkl<a> dislocations in zirconium, with basal slip being easier to activate, 
at least in pure zirconium.  
\Abinitio{} calculations \cite{Chaari2014} concluded that these two slip systems 
are indeed intertwined as the same energy barrier is obtained
for a straight screw dislocation gliding either in a basal or a pyramidal plane. 
The dislocation, which remains dissociated in a prismatic plane, glides in both cases
through the conservative motion of the stacking fault ribbon perpendicular to the dissociation plane,
without any constriction of the dissociated core.
Most importantly, these \abinitio{} calculations have shown that 
basal glide is a combination of prismatic and pyramidal glide. 
Basal slip should therefore be composite. 
But this is not compatible with experimental results described above, 
in particular the observation of slip traces fully resolved in the basal planes \cite{Gong2015,Caillard2018}.
As underlined by Caillard \etal{} \cite{Caillard2018}, glide of \hkl<a> dislocations 
in basal planes should be an elementary slip system in zirconium.
It it the purpose of this article to look how the mechanism 
given by \abinitio{} calculations for secondary slip systems of \hkl<a> dislocations
in zirconium, in particular basal glide,
can be reconciled with the various experimental observations.
As it appears necessary to go beyond the motion of an infinite straight screw dislocation,
we use molecular dynamics simulations to study glide of a screw dislocation in zirconium,
thus allowing for localised thermally activated events like nucleation of kink pairs. 
The obtained glide mechanism is then analysed in more details 
and in a larger stress range with energy barrier calculations. 
Finally, we discuss how results of atomic simulations match with experiments.

\section{Molecular dynamics simulations}

\subsection{Methods}

All atomic simulations are carried out with the embedded atom method (EAM)
potential developed by Mendelev and Ackland (potential \#3) \cite{Mendelev2007}.
This potential predicts the correct ground state of the \hkl<a> screw dislocation
\cite{Clouet2012} with a core dissociated in the prismatic plane 
in two $\hkl<a>/2$ partial dislocations, in agreement with \abinitio{} calculations
\cite{Clouet2012,Chaari2014,Clouet2015}.
The infinite stable stacking fault found in the prismatic plane with this potential
has a slightly different atomic structure, in particular a different fault vector,
than the \abinitio{} one, 
but this does not result then in any significant dissimilarity 
for the relaxed configuration of the dislocation core \cite{Clouet2012}. 
A metastable configuration where the screw dislocation dissociates in two Shockley partials in the basal plane
is also obtained with this potential, while \abinitio{} calculations show that such a configuration
should be unstable \cite{Clouet2012}.  But, as this spurious configuration has a much higher energy 
than the prismatic ground state ($\Delta E=62$\,meV\,{\AA}$^{-1}$), it takes no part during dislocation motion.
This potential actually leads to the same mechanism for basal glide of a straight screw dislocation
as \abinitio{} calculations: 
basal glide operates through displacement of the prismatic stacking fault
perpendicular to its habit plane leading to a metastable configuration, halfway through the migration,
where the dislocation is partially spread in a pyramidal plane \cite{Chaari2014,Chaari2014a}. 
During this elementary slip event, the dislocation trajectory can be decomposed into three steps: 
the dislocation glides first in a prismatic plane, then in the pyramidal plane 
and finally again in a prismatic plane, with an energy barrier opposing dislocation motion 
only in the pyramidal part of the trajectory.

The simulation box is oriented with $x$ axis along the dislocation direction ($x \parallel \hkl[11-20]$), 
$y$ along the glide direction in the basal plane ($y \parallel \hkl[1-100]$), 
and $z$ orthogonal to the basal glide plane ($z \parallel \hkl[0001]$).
It is periodic along $x$ and $y$ directions, with controlled surfaces along the $z$ direction.
The corresponding sizes are $l_x=156$, $l_y=140$ and $l_z=1340$\,\AA, leading to a cell containing 1.2 million atoms.
As the screw dislocation can easily glide in its prismatic plane in the $z$ direction, 
large size $l_z$ is needed to allow thermal fluctuations of the dislocation position in the prismatic plane
while preventing it to get out of the slab.
The dislocation length $l_x = 48b$ is large enough to allow for the nucleation of kink pairs 
in the considered stress range.

The screw dislocation is introduced in the middle of the simulation box using the displacement field predicted by elasticity theory.
Upon relaxation, the dislocation spontaneously dissociates in two partials in the prismatic plane. 
The temperature is initialised to the target temperature. After an initial thermalisation stage, 
molecular dynamics (MD) simulations are performed with a 2\,fs time step in the $NVE$ ensemble without any thermostat: 
despite the work of the applied stress, the cell is large enough for the temperature increase 
to be negligible during the whole dynamics. 
We use stress controlled simulations \cite{Rodney2007}: 
an average force $\pm F_x$ is imposed on the atoms in the outer layers close to the two $z$-surfaces 
to apply a shear stress $\tau=\tau_{xz}$. 
This shear stress leads to a Peach-Koehler force on the dislocation resolved in the $y$-direction,
thus promoting basal glide, without any resolved shear stress in the prismatic plane to prevent easy glide in this plane.
All simulations are carried out with \lammps{} \cite{Plimpton1995}
for applied stresses in the range 600 - 1200\,MPa 
and for temperatures in the range 300 - 900\,K.

\subsection{Dislocation velocity}

\begin{figure}[!b]
	\begin{center}
		\includegraphics[width=0.8\linewidth]{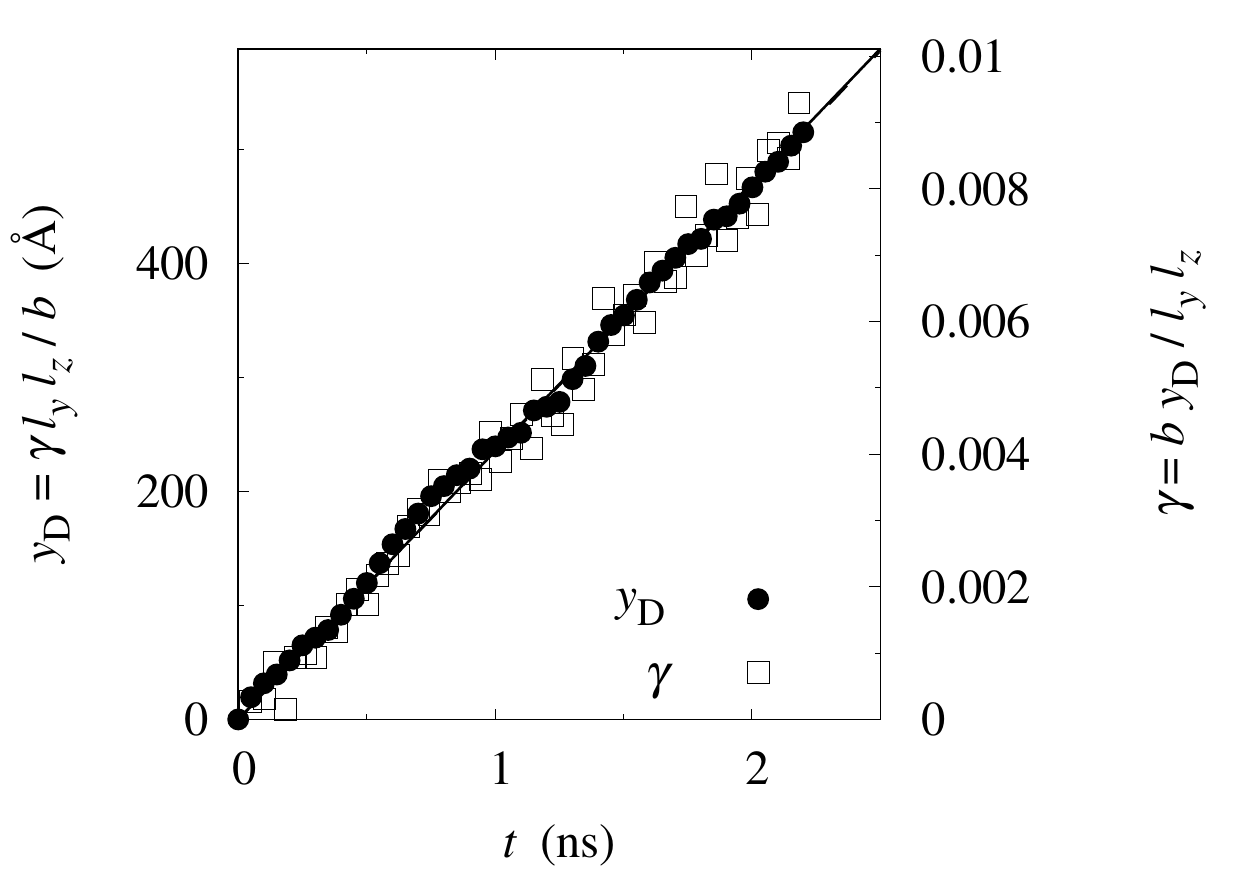}
	\end{center}
	\caption{Evolution with time $t$ of the dislocation position $y_{\rm D}$ 
	given by DXA and of the shear strain $\gamma$ deduced from the displacement 
	of the outer layers during an MD simulation with an applied stress $\tau=900$\,MPa 
	and a temperature $T=600$\,K.  Orowan relation imposes 
	$\gamma(t)-\gamma(0) = b \left[ y_{\rm D}(t) - y_{\rm D}(0) \right] / l_y l_z$.}
	\label{fig:MD_dislo_position}
\end{figure}

All MD simulations lead to the same glide mechanism, whatever the applied stress
and the temperature.  Basal glide operates 
through the nucleation and the propagation of kink pairs,
with the dislocation moving one Peierls valley at a time. 
We extract the dislocation position from these simulations using DXA algorithm \cite{Stukowski2012a}
implemented in \ovito{} \cite{Stukowski2010a}.
As shown in Fig. \ref{fig:MD_dislo_position}, the velocity $v_{\rm D}$ deduced from the linear variation of this dislocation position 
agrees with the average strain rate $\dot{\gamma}$ deduced from the relative displacement 
of the two outer layers used to control the stress, through Orowan relation 
$\dot{\gamma} = b v_{\rm D} / l_y l_z$.

\begin{figure}[!b]
	\begin{center}
		\includegraphics[width=0.99\linewidth]{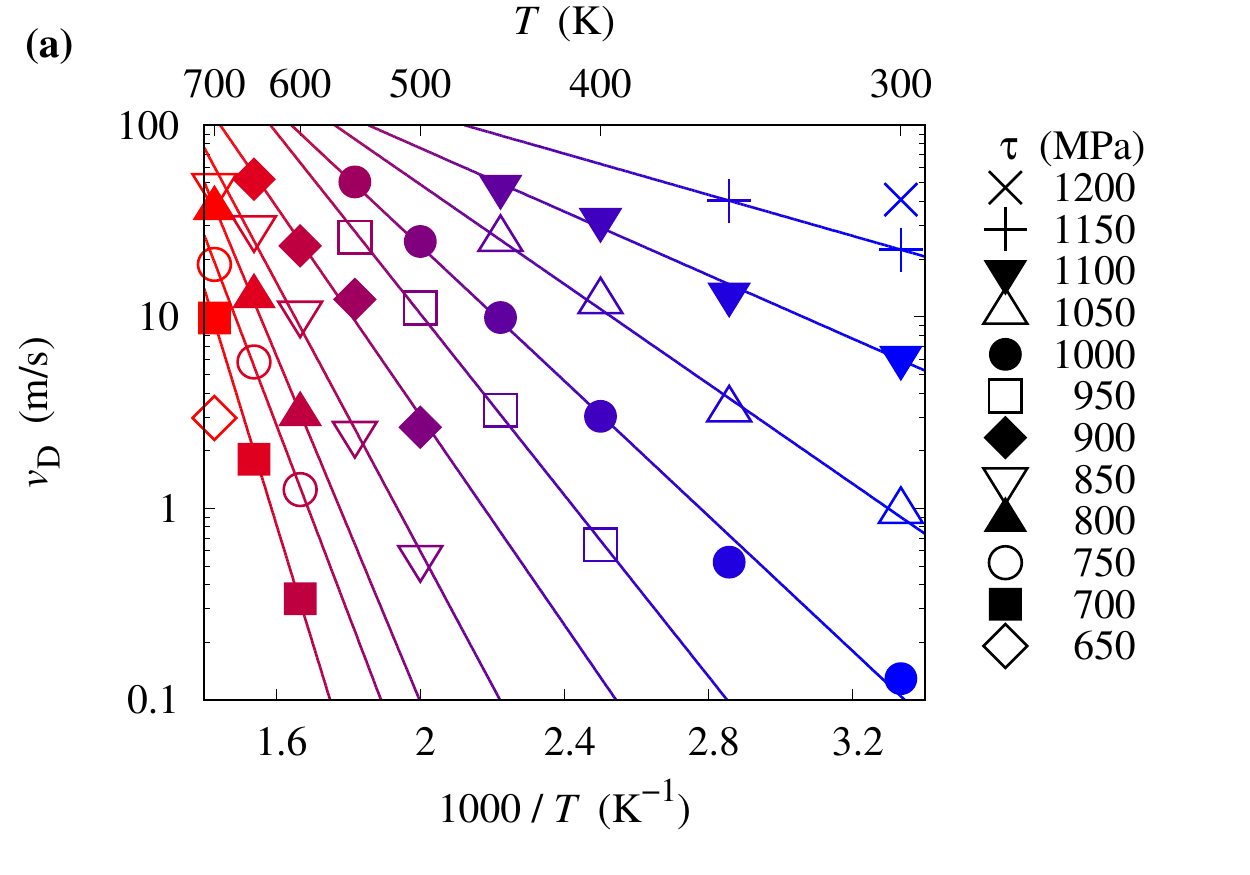}
		\includegraphics[width=0.99\linewidth]{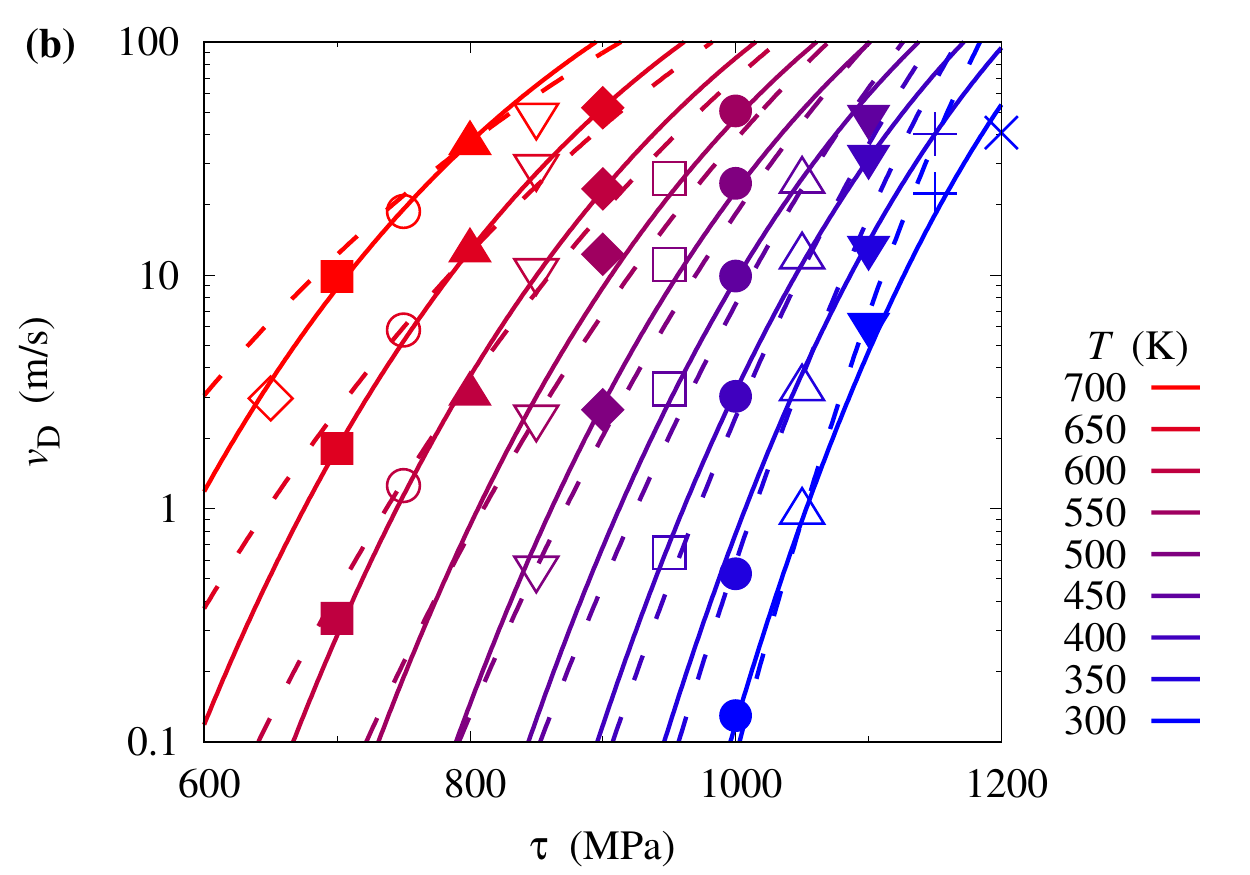}
	\end{center}
	\caption{Dislocation velocity as a function of the temperature $T$ 
	and the applied stress $\tau$. 
	Symbols are results of MD simulations. 
	Lines in (a) are the fits of Arrhenius equation used to obtain, for each stress $\tau$,
	the activation enthalpies and velocity prefactors shown in Fig. \ref{fig:MD_activation}.
	Lines in (b) are the velocities given by Eq. \ref{eq:velocity_final} 
	with parameters given either by fitting separately activation enthalpies and velocity prefactors 
	(dashed lines) or by fitting directly dislocation velocities (solid lines).
	}
	\label{fig:MD_velocity}
\end{figure}

\begin{figure}[!b]
	\begin{center}
		\includegraphics[width=0.99\linewidth]{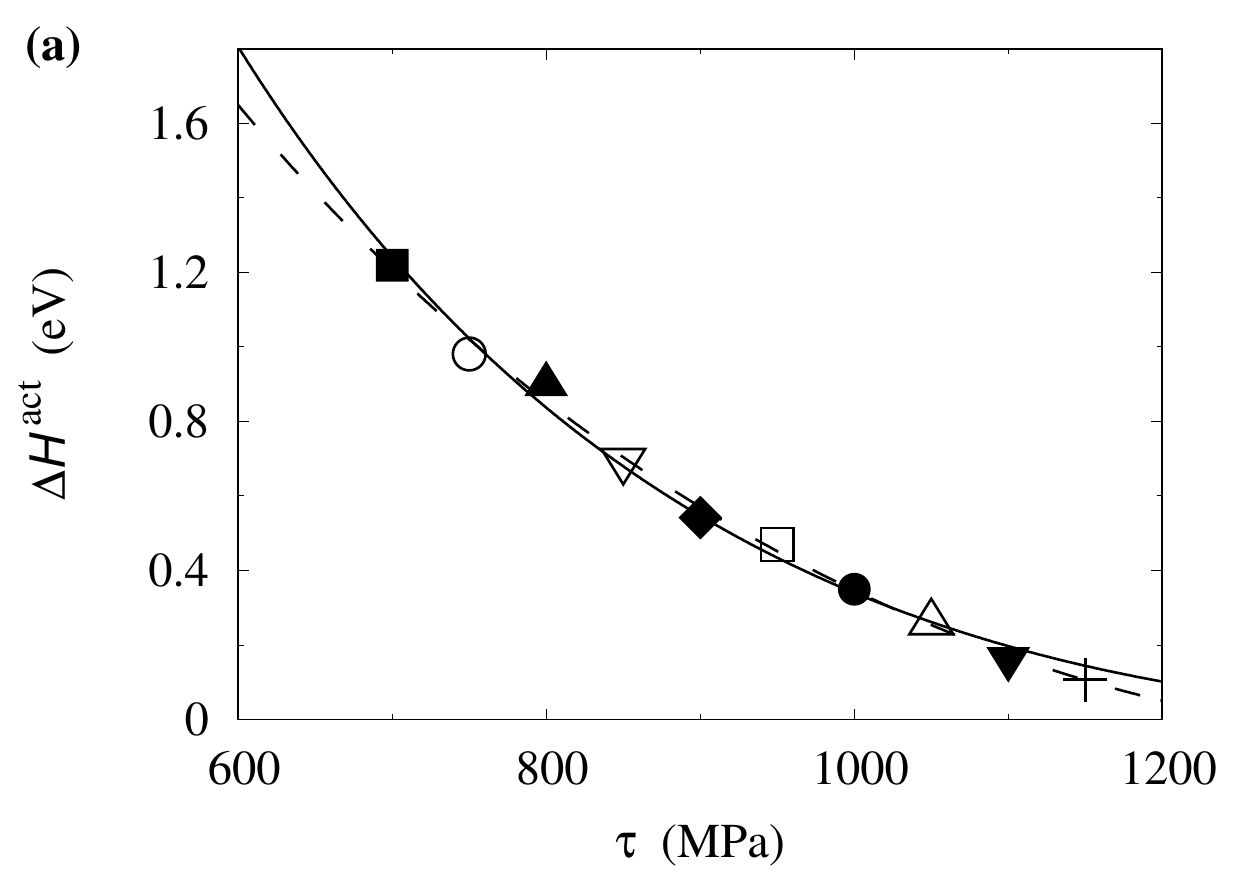}
		\includegraphics[width=0.99\linewidth]{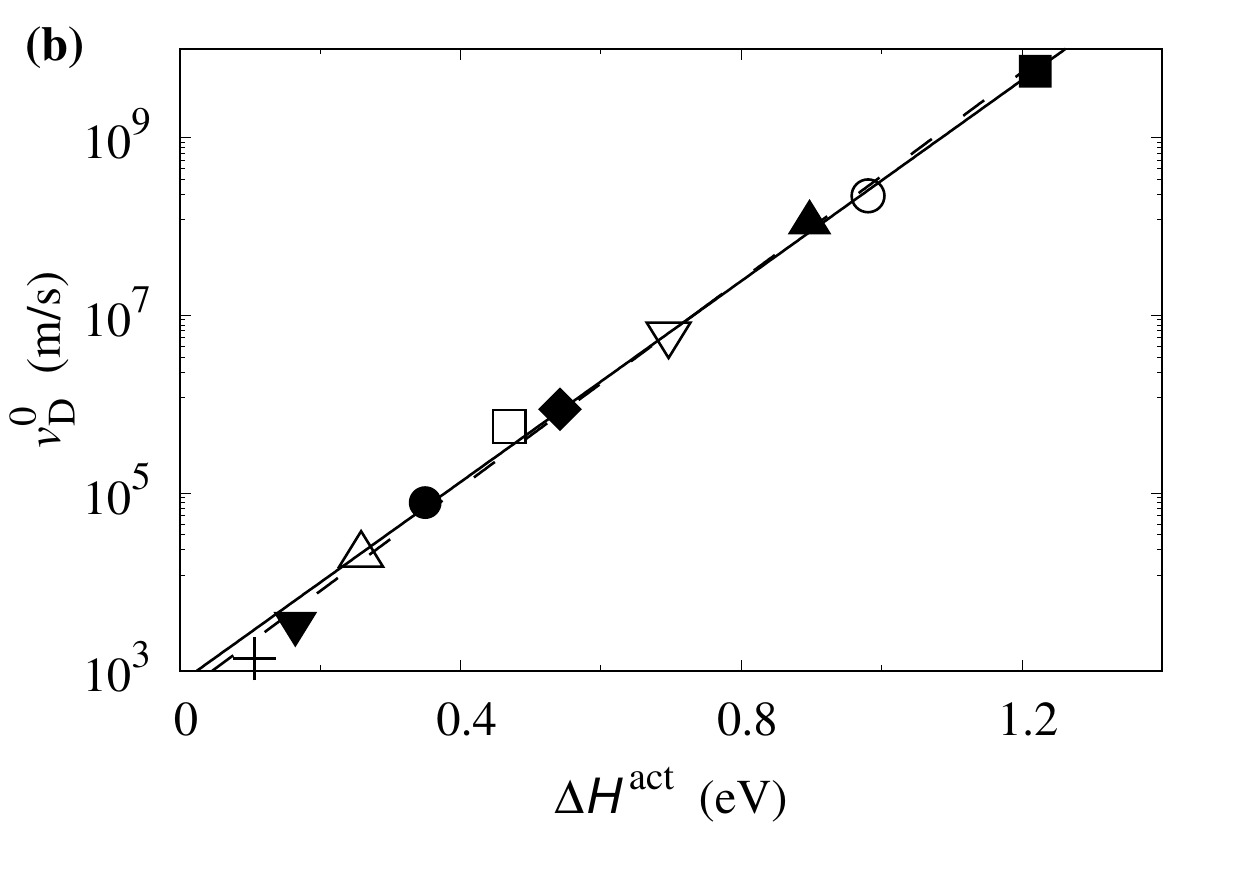}
	\end{center}
	\caption{Parameters defining thermally activated law for dislocation basal glide:
	(a) activation enthalpy $\Delta H^{\rm act}$ as a function of the applied stress $\tau$
	and (b) velocity prefactor $v^0_{\rm D}$ as a function of the activation enthalpy.
	Symbols correspond to results of MD simulations and are obtained through the fits 
	of Arrhenius equation shown in Fig. \ref{fig:MD_velocity}a.
	Lines are analytical expressions corresponding to (a) Kocks law (Eq. \ref{eq:Kocks})
	and (b) Meyer-Neldel rule (Eq. \ref{eq:prefactor_log}). 
	The dashed lines are obtained through separate fits of the MD activation enthalpy 
	and velocity prefactors,
	whereas solid lines are obtained through a global fit
	of all dislocation velocities measured in MD simulations 
	directly by Eq. \ref{eq:velocity_final} (Fig. \ref{fig:MD_velocity}b).}
	\label{fig:MD_activation}
\end{figure}

Dislocation velocities obtained from MD simulations for different applied stresses $\tau$ 
and different temperatures $T$ are shown with symbols in Fig. \ref{fig:MD_velocity}.
For a given stress $\tau$, the variation with temperature of the dislocation velocity 
follows a thermally activated law:
\begin{equation}
	v_{\rm D}(\tau,T) = v^0_{\rm D}(\tau) \exp{\left[ - \Delta H^{\rm act}(\tau)/kT \right]}.
	\label{eq:velocity}
\end{equation}
We use the linear variations observed in the Arrhenius plot (Fig. \ref{fig:MD_velocity}a) 
to extract, for each applied stress, the activation enthalpy $\Delta H^{\rm act}(\tau)$
and the velocity prefactor $v^0_{\rm D}(\tau)$. 
The obtained quantities are shown with symbols in Fig. \ref{fig:MD_activation}.

\begin{table}[!tb]
	\caption{Parameters of Kocks law (Eq. \ref{eq:Kocks})
	and of the Meyer-Neldel compensation rule (Eq. \ref{eq:prefactor})
	deduced from MD simulations
	either by separate fits of the activation enthalpy
	and of the velocity prefactor (dashed lines in Fig. \ref{fig:MD_velocity}b and \ref{fig:MD_activation})
	or directly by a global fit of the dislocation velocities (solid lines in Fig. \ref{fig:MD_velocity}b and \ref{fig:MD_activation}).}
	\label{tab:mobility_para}
	\centering
	\begin{tabular}{lcc}
		\hline
					& Separate fit	&  Global fit	\\
		\hline
		$H^0$ (eV)		& 67.8		& 101	\\
		$\tau_{\rm P}$ (MPa)	& 1280		& 1540	\\
		$p$			& 0.100		& 0.201	\\
		$q$			& 1.42		& 2.30	\\
		$T_{\rm m}$ (K)		& 863		& 891	\\
		$\nu$ (THz)		& 0.040		& 0.054	\\
		\hline
	\end{tabular}
\end{table}

Variations of the activation enthalpy are well described by a Kocks law \cite{Kocks1975}:
\begin{equation}
	\Delta H^{\rm act}(\tau) = H^0 \left[ 1 - \left( \frac{\tau}{\tau_{\rm P}} \right)^p \right]^q.
	\label{eq:Kocks}
\end{equation}
The resulting fit is shown with a dashed line in Fig. \ref{fig:MD_activation}a 
and the corresponding four parameters are given in table \ref{tab:mobility_para} 
(column ``Separate fit''). 
As it will be shown below, these parameters are very sensitive to the way the fit is performed 
and the obtained Kocks law should be mainly understood as a way to interpolate
MD results in the corresponding stress range ($700 \leq \tau \leq 1150$\,MPa), 
with a poor predictive ability out of this fitting range.

A linear relation between the logarithm of the velocity prefactor $\log{\left[ v^0_{\rm D}(\tau) \right]}$
and the activation enthalpy $\Delta H^{\rm act}(\tau)$ is observed (Fig. \ref{fig:MD_activation}b). 
This linear relation can be rationalised by writing this velocity prefactor 
as the product of a constant term which does not depend of the applied stress 
and of an entropic contribution.
Assuming a length dependent regime for dislocation mobility \cite{Caillard2003},
this velocity prefactor is written
\begin{equation}
	v^0_{\rm D}(\tau) = \frac{l_{\rm D}}{b} \lambda_P \, \nu \exp{\left[ \Delta S^{\rm act}(\tau) / k \right]}
	\label{eq:prefactor}
\end{equation}
where $l_{\rm D}$ is the length of the dislocation line 
and $l_{\rm D}/b$ therefore the number of possible nucleation sites for a kink pair,
$\lambda_{\rm P}=a\sqrt{3}/2$ the length between two Peierls valleys in the basal glide direction,
$\nu$ the fundamental attempt frequency,
and $\Delta S^{\rm act}(\tau)$ the activation entropy.
Following Meyer-Neldel compensation rule \cite{Meyer1937}, we assume that activation entropy is proportional to activation enthalpy,
$\Delta S^{\rm act}(\tau) = \Delta H^{\rm act}(\tau) / T_{\rm m}$,
thus leading to the linear relation
\begin{equation}
	\log{\left[ v^0_{\rm D}(\tau) \right]} = \log{\left[ \frac{l_{\rm D}}{b} \lambda_P \, \nu \right]}
	+ \frac{\Delta H^{\rm act}(\tau)}{k T_{\rm m}},
	\label{eq:prefactor_log}
\end{equation}
in agreement with results of MD simulations (Fig. \ref{fig:MD_activation}b).
The fit of the velocity prefactor $v^0_{\rm D}(\tau)$ 
as a function of the activation enthalpy $\Delta H^{\rm act}(\tau)$ 
(dashed line in Fig. \ref{fig:MD_activation}b) 
leads to the parameters $\nu$ and $T_{\rm m}$ shown in table \ref{tab:mobility_para} (column ``Separate fit'').
As expected \cite{Saroukhani2016,EstebanManzanares2020}, the attempt frequency is about one hundredth of the Debye frequency (6\,THz \cite{Kittel1996})
and the scaling temperature $T_{\rm m}$ is close to the melting temperature 
(a temperature of 1363\,K is obtained with this EAM potential for the melting of the hcp phase
\cite{Mendelev2007}).

At the end, we obtain an analytical expression describing the dislocation velocity
as a function of the applied stress and of the temperature,
\begin{multline}
	v_{\rm D}(\tau,T) = \frac{l_{\rm D}}{b} \lambda_P \, \nu \\
	\exp{\left\{ - \frac{H^0}{kT} \left[ 1 - \left( \frac{\tau}{\tau_{\rm P}} \right)^p \right]^q
		\left[ 1 - \frac{T}{T_{\rm m}} \right] \right\}},
	\label{eq:velocity_final}
\end{multline}
with all unknown parameters obtained from the fits of the activation enthalpy 
and of the velocity prefactor.
The dislocation velocities extracted from MD simulations are compared to this analytical expression 
in Fig. \ref{fig:MD_velocity}b (dashed line).
A reasonable agreement is obtained, except for the lowest (650\,MPa) and the highest stresses (1150 and 1200\,MPa), 
for which MD simulations have been performed for too few temperatures, 
thus preventing a precise determination of the activation enthalpy and of the velocity prefactors.
To improve the agreement, we fit directly Eq. \ref{eq:velocity_final} to the dislocation velocities
obtained by MD for all temperatures and applied stresses, 
using the four parameters of the Kocks law and the two parameters of the Meyer-Neldel rule 
as variables (global fit). 
This leads to the solid lines shown in Fig. \ref{fig:MD_velocity}b which perfectly match
the results of MD simulations for all considered stresses and temperatures.
The corresponding variations of the activation enthalpy and of the velocity prefactors (solid lines in Fig. \ref{fig:MD_activation})
are not really different from the ones obtained with the previous fits (dashed lines in Fig. \ref{fig:MD_activation}) 
in the stress range $700\leq\tau\leq1100$\,MPa, but lead to a slightly different extrapolation out of this range.
This is mainly true for the activation enthalpy where the global fit leads to different parameters 
for the Kocks law (Tab. \ref{tab:mobility_para}).
As a consequence, the obtained analytical law can describe dislocation velocity for stresses higher than 600\,MPa
where it has been possible to perform MD simulations, 
but its use for extrapolation at lower stresses appears hazardous.
Finally, is is worth pointing that the Peierls stress obtained by this fit of MD results, 
$\tau_{\rm P}=1.54$\,GPa, is close from the value previously obtained with the same EAM potential
(1.79\,GPa) by looking for the the stress cancelling the energy barrier for pyramidal and basal glide 
for a straight screw dislocation \cite{Chaari2014,Chaari2014a},
\ie{} without considering the nucleation and propagation of kink pairs.

\subsection{Dislocation trajectory}

\begin{figure}[!b]
	\begin{center}
		\includegraphics[width=0.8\linewidth]{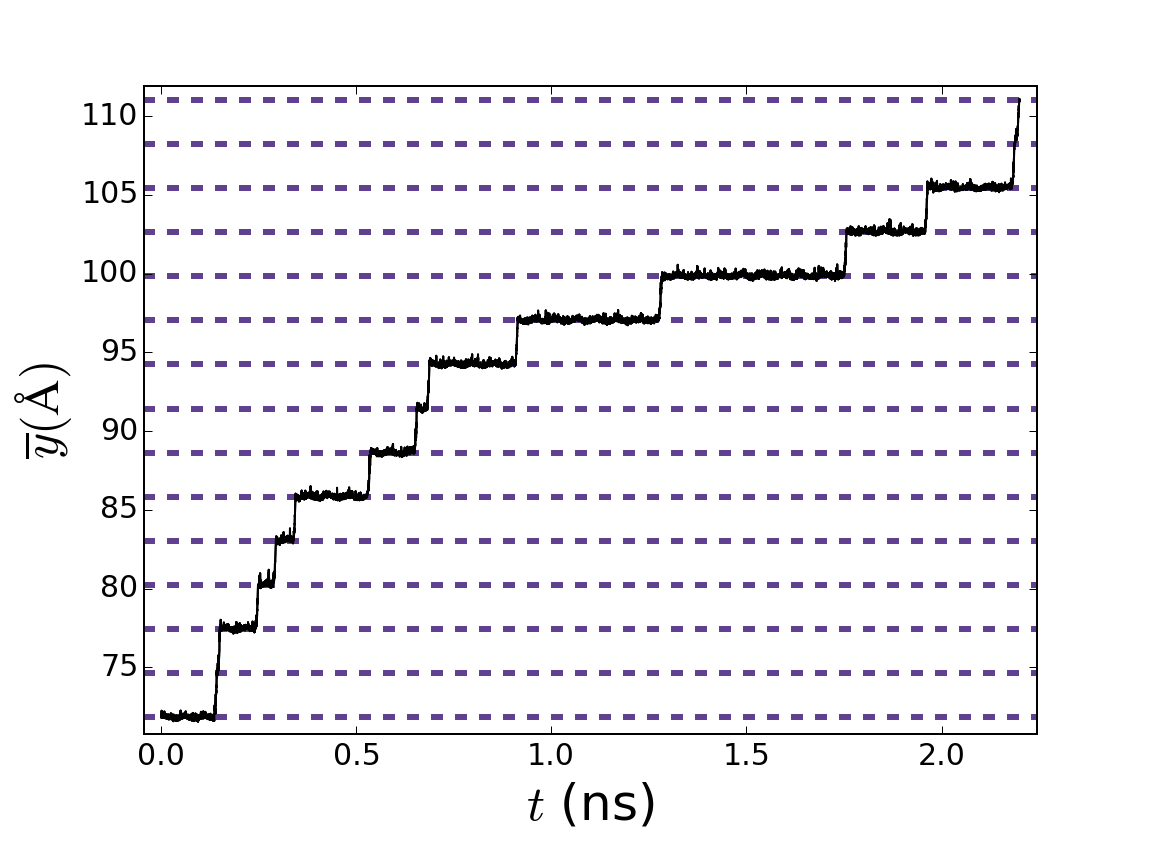}
		\includegraphics[width=0.8\linewidth]{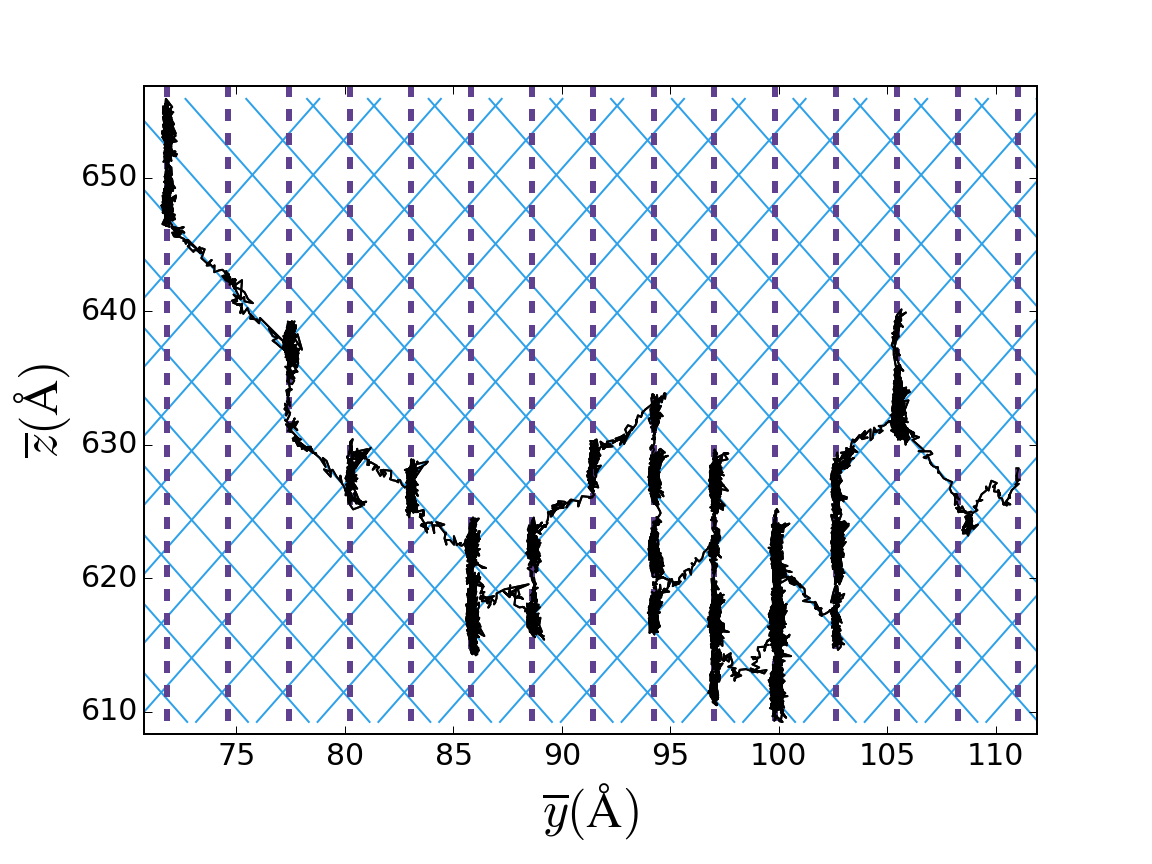}
	\end{center}
	\caption{Dislocation glide in the basal plane at 600\,K for an applied stress $\tau=750$\,MPa.
	The dislocation average position $\bar{y}$ in the basal plane is represented 
	as a function of time $t$ in the upper figure,
	whereas the dislocation trajectory, with $\bar{z}$ the average dislocation position in the prismatic plane,
	is shown in the lower figure. 
	The positions of the prismatic and pyramidal planes corresponding to the bottom of the Peierls valleys
	are respectively indicated with dashed purple and continuous blue lines.}
	\label{fig:MD_600K_750MPa}
\end{figure}

Looking in more details to the dislocation motion (Fig. \ref{fig:MD_600K_750MPa}),
one sees that it can be decomposed in jump events between two Peierls valleys in the basal plane ($y$ direction)
separated by waiting times where the dislocation fluctuates in its prismatic habit plane ($z$ position).
As there is no resolved shear stress in the prismatic plane, this fluctuation of the $z$ position
corresponds to a Brownian motion of the dislocation in this prismatic plane where it can easily glide 
with almost no energy barrier \cite{Clouet2012}.
The jumps correspond to the nucleation and propagation of a kink pair, as described by the thermally activated
mobility law (Eq. \ref{eq:velocity_final}).
The trajectory followed by the dislocation during these jumps does not lie in the basal plane,
but in a first order pyramidal plane (Fig. \ref{fig:MD_600K_750MPa}).
For each jump, the dislocation can glide in one of the two available pyramidal planes.
As the only non-null component of the applied stress is $\tau_{xz}$, 
the two pyramidal planes see exactly the same resolved shear stress, thus leading to an equal probability 
for the dislocation to transit by one of these two planes. 
As a consequence, the average glide plane coincides with the basal plane. 
But this is clearly a consequence of our simulation setup where the basal plane 
is the maximum resolved shear stress plane (MRSSP).
With such a glide mechanism where the dislocation glides from one Peierls valleys to the other
by transiting through a pyramidal plane, the macroscopic glide plane should correspond to the MRSSP
which would be the basal plane only for very specific loading conditions where the other shear stress component
$\tau_{yz}$ is equal to zero. 
This is in contradiction with experiments \cite{Caillard2018} which reports basal slip 
as an elementary slip system.
As it will be shown in the next section, the difference may be ascribed to higher applied stresses
in molecular dynamics simulations than in experiments.

\section{From high to low stress}

We now perform nudged elastic band (NEB) calculations \cite{Henkelman2000a}
to study 
the mechanisms controlling basal glide
in a larger stress range than what can be simulated with MD, in particular at lower stresses.  
These calculations allow the determination of the minimum energy path
for the screw dislocation gliding in a basal plane between two neighbouring Peierls valleys. 
The simulation setup is the same as for the MD simulations, 
with only a smaller cell length in the \hkl[0001] direction perpendicular to the basal glide plane
($l_z=278$\,\AA).
To ensure that the dislocation is not simply moving as a whole 
but is gliding through the nucleation and propagation of kink pairs,
as observed in MD simulations, 
the images defining the starting path are initialised with segments lying in one of the two Peierls valleys,
with the length of the part in the arrival valley being proportional to the image index in the path.

\subsection{Enthalpy barriers for kink nucleation}

\begin{figure}[!b]
	\begin{center}
		\includegraphics[width=0.99\linewidth]{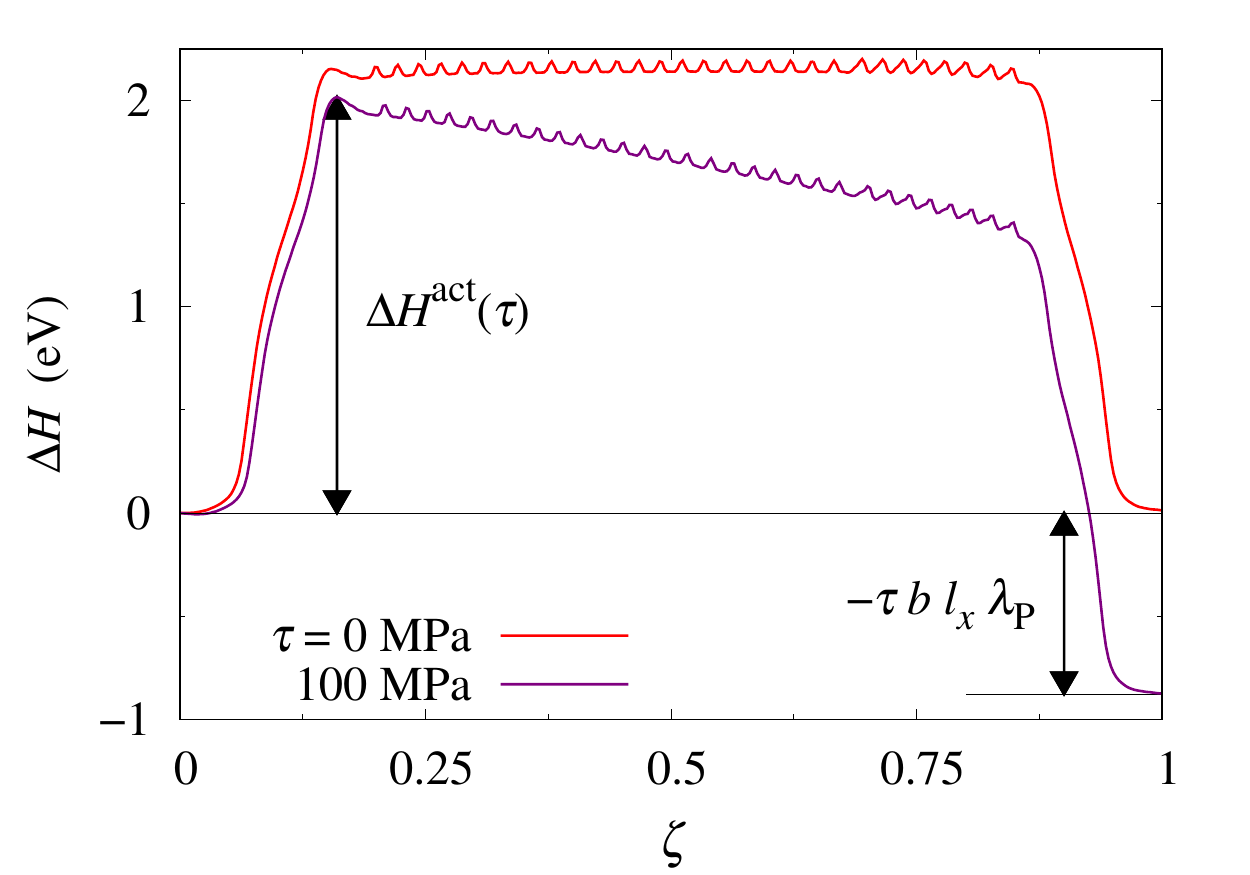}
	\end{center}
	\caption{Enthalpy barriers $\Delta H(\zeta)$ obtained by NEB for basal glide of a screw dislocation 
	for two different applied stresses $\tau$.
	$\zeta$ is the NEB reaction coordinate along the minimum energy path.}
	\label{fig:NEB_barrier}
\end{figure}

Some enthalpy barriers obtained for basal glide are shown in Fig. \ref{fig:NEB_barrier}.
In the absence of an applied stress, this barrier is the superposition of an almost constant plateau at 2.2\,eV
corresponding to the formation energy of a kink pair and small undulations due to the migration of the kinks along the dislocation line,
with a migration energy equal to 0.05\,eV. 
Kink migration appears thus much easier than kink pair nucleation. 
Besides, the formation energy is almost independent from the size of the kink pair, 
showing that the elastic interaction between kinks can be neglected.

When a positive stress $\tau$ is applied, the enthalpy barrier is shifted with a slope
corresponding to the work of the Peach-Koehler forces (Fig. \ref{fig:NEB_barrier}), 
\ie{} $-\tau\,b\,l_x\,\lambda_{\rm P}$ for the final position ($\zeta=1$)
where the whole dislocation of length $l_x$ has glided a distance $\lambda_{\rm P}$.
The highest enthalpy is met at the beginning of the path when the kink pair is nucleated. 
This defines the activation enthalpy $\Delta H^{\rm act}(\tau)$
which controls the basal mobility of the screw dislocation for an applied stress $\tau$.

\begin{figure}[!bt]
	\begin{center}
		\includegraphics[width=0.99\linewidth]{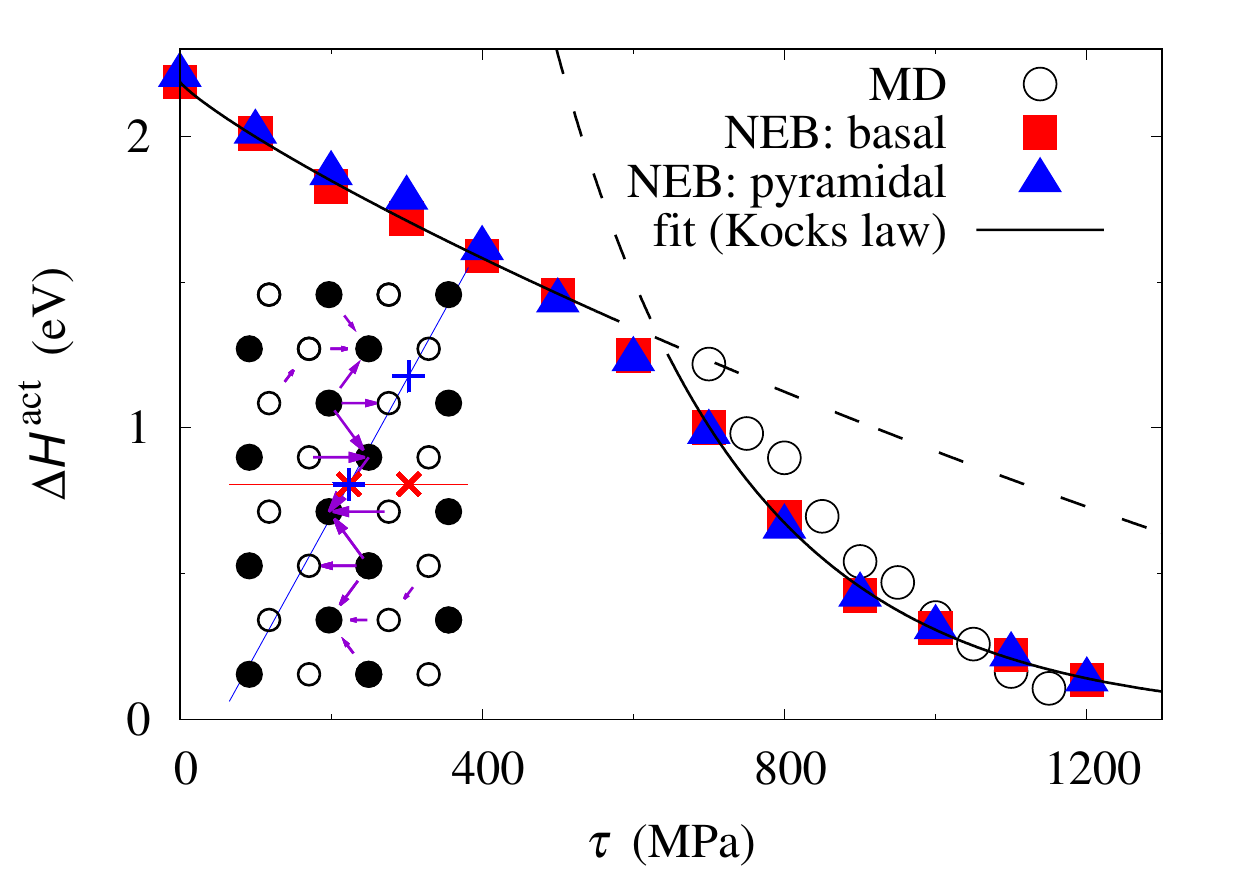}
	\end{center}
	\caption{Variation with the applied stress $\tau$ of the activation enthalpy $\Delta H^{\rm act}$
	for dislocation glide. 
	NEB calculations have been performed for basal and pyramidal glide, 
	with the respective initial and final dislocation positions indicated with blue and red crosses
	on the differential displacement map shown in the inset.
	The obtained enthalpies are compared to MD results 
	and are used to fit Kocks laws in two different stress regimes.}
	\label{fig:activation_NEB}
\end{figure}

We calculate this activation enthalpy $\Delta H^{\rm act}(\tau)$
for different applied stresses $\tau$.
Results are shown with red squares in Fig. \ref{fig:activation_NEB}.
Activation enthalpies given by NEB calculations can be compared with the ones
previously extracted from MD simulations when the applied stress $\tau$ is higher than 700\,MPa.
A good agreement is obtained between both quantities, 
further confirming that the glide mechanism given by NEB 
is the same as the one controlling dislocation mobility in MD simulations.
The small differences appearing in Fig. \ref{fig:activation_NEB} 
may arise from variations with temperature: MD simulations are performed at finite temperatures
whereas NEB calculations are for 0\,K.

Looking at the variations with the applied stress of the activation enthalpy,
a discontinuity can be seen around 600\,MPa.  
Whereas the activation enthalpy decreases almost linearly with the stress
up to 600\,MPa, a faster decrease is observed above (Fig. \ref{fig:activation_NEB}).
As a consequence, it is not possible to fit the activation enthalpy with a single Kocks law 
in the whole stress range, but two laws with different parameters are needed. 
This suggests a change in the mechanism controlling basal glide, 
with a low- and a high-stress regime which become competitive at 600\,MPa.
As it will be shown below, this is confirmed by analysing the shape of the critical nucleus
associated with this activation enthalpy at different stresses.

\subsection{Critical nucleus and activation volume}

\begin{figure}[!b]
	\begin{center}
		\includegraphics[width=0.99\linewidth]{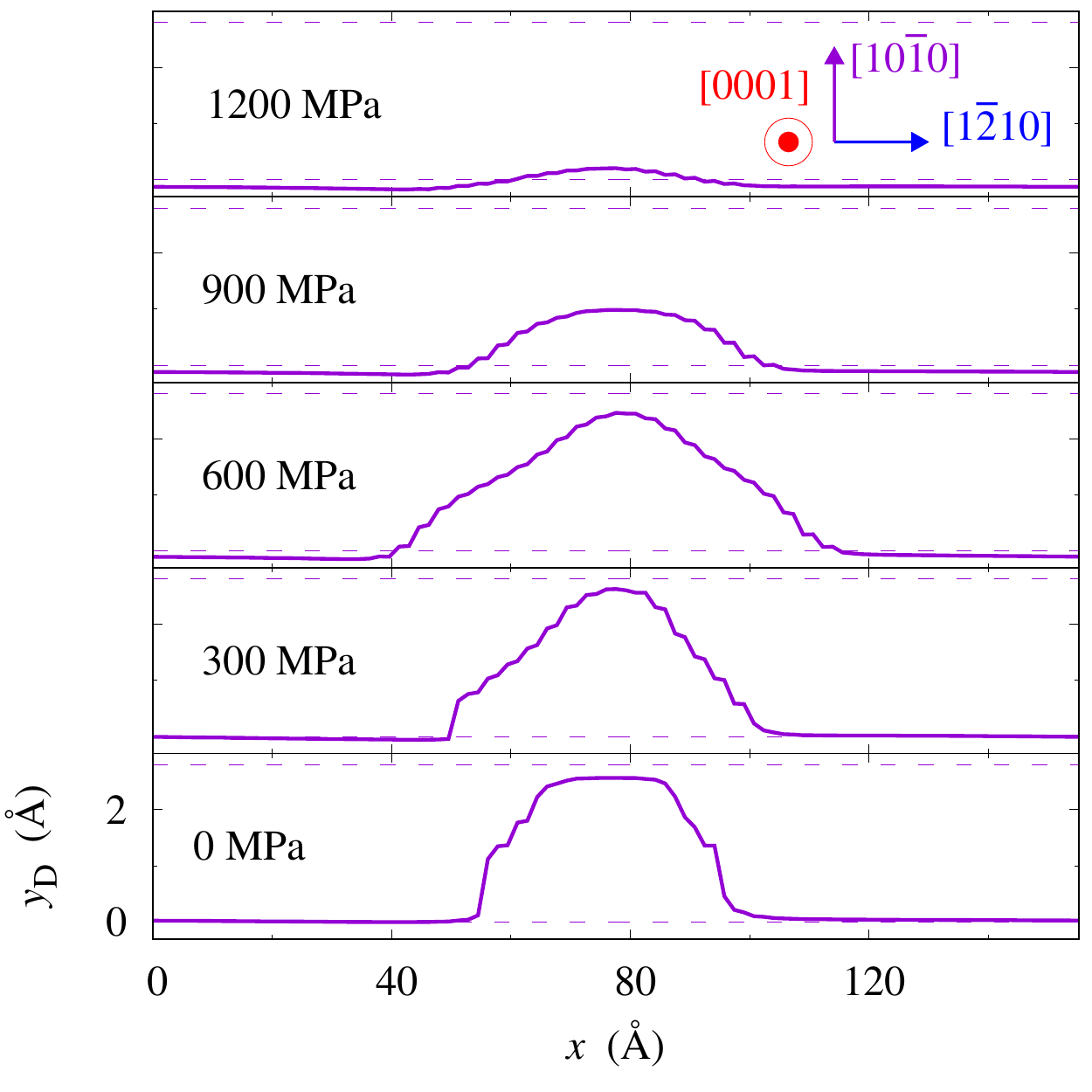}
	\end{center}
	\caption{Projected shape in the basal plane of the critical nucleus 
	corresponding to basal or pyramidal glide of the screw dislocation 
	for different applied stresses $\tau$.
	The dashed lines indicate the bottom of the Peierls valleys.}
	\label{fig:critical_kink}
\end{figure}

The shape of the critical nucleus is obtained by considering the atomic configuration
with the highest enthalpy along the minimum energy path for a given applied stress $\tau$. 
The disregistry $D(x,y)$ created by the kinked dislocation is first extracted by taking the displacement difference
between atoms located in the plane just above and just below the basal glide plane
for every possible position $(x,y)$. 
The solution given by Peierls-Nabarro model is then fitted to the obtained disregistry \cite{Clouet2018},
in strips of width $\Delta x=2b$ along the line, thus allowing the definition of the dislocation position $y_{\rm D}(x)$.
The dislocation profiles obtained for the critical nucleus are shows in Fig. \ref{fig:critical_kink}
for different applied stresses.  
Similar shapes are obtained when using DXA or when identifying the dislocation position 
with the average positions of the most energetic atoms
in each $\Delta x$ strip.

For applied stresses lower than 600\,MPa, the critical nucleus extends from one Peierls valley
to the next valley with its extension in the initial valley becoming wider
when the stress increases.  
Above 600\,MPa, the critical nucleus does not jump over the barrier completely up to the next valley and
the kink height decreases with an increasing stress.
A change of the critical kink pair shape therefore occurs around 600\,MPa, 
at the transition between the low- and high-stress regimes previously seen for the activation enthalpy.

\begin{figure}[!bt]
	\begin{center}
		\includegraphics[width=0.99\linewidth]{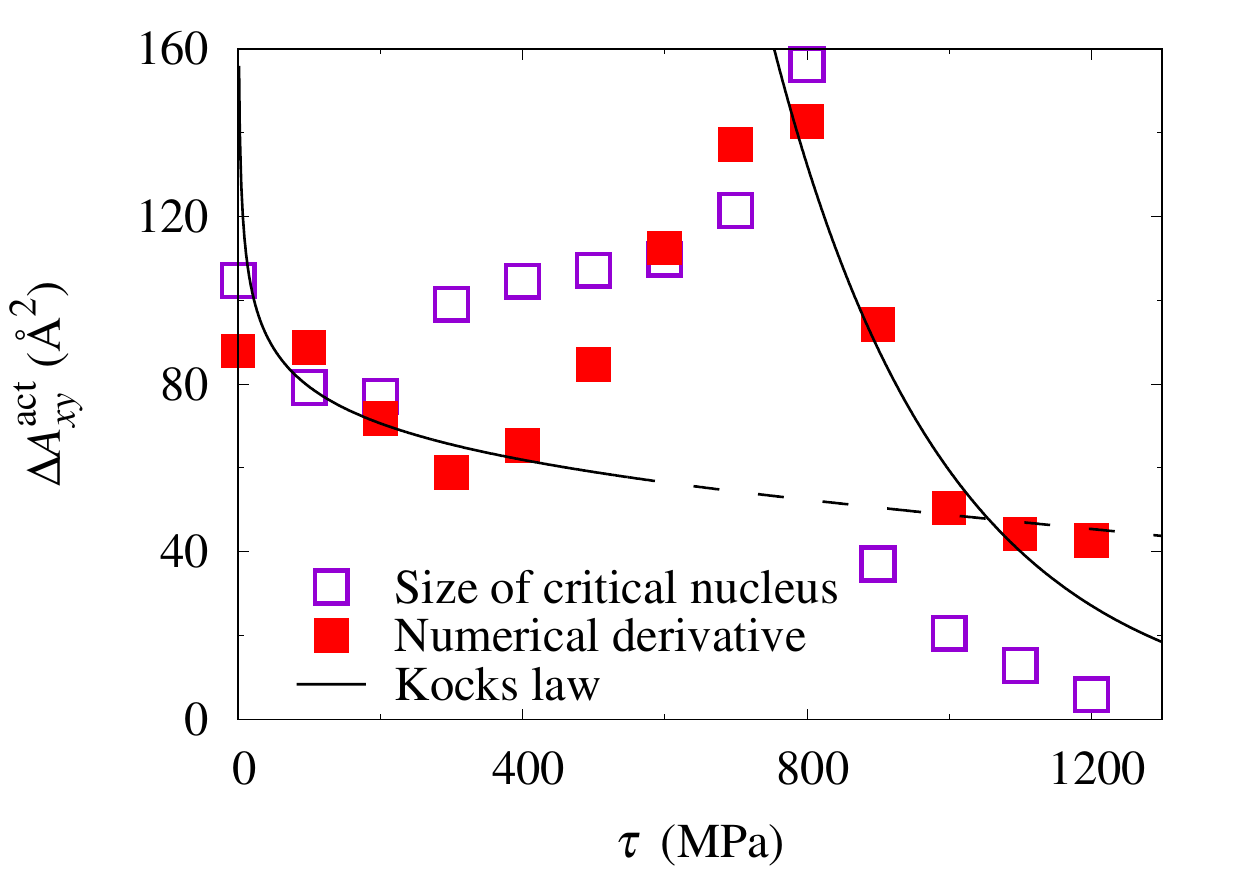}
	\end{center}
	\caption{Variation with the applied stress $\tau$ of the activation area $\Delta A^{\rm act}_{xy}$
	for basal glide.
	The activation area is calculated either by considering the swept area of the critical nucleus (Fig. \ref{fig:critical_kink}), 
	the numerical derivative of the activation enthalpy $\Delta H^{\rm act}$ (Fig. \ref{fig:activation_NEB})
	or the analytical derivative of the corresponding Kocks laws.}
	\label{fig:activation_volume}
\end{figure}

This change of the critical nucleus shape with the applied stress
is consistent with predictions of simple models using a line tension approximation
to describe the dislocation and integrating its interaction with the crystal 
trough a 1D Peierls potential 
(see chap. 4.2 in Ref. \cite{Caillard2003} for a thorough discussion on these models):
at high stress, the critical nucleus corresponds to a dislocation bulge,
whereas it converges towards two well-separated kinks at low stress.
The transition between these two regimes can be estimated by considering 
that the bulge correctly describes the critical nucleus as long as 
its head does not reach the bottom of the next Peierls valley, 
leading to a minimum stress $\tau^* = \Delta E_{\rm P} / b\lambda_{\rm P}$, 
with $\Delta E_{\rm P}$ the height of the Peierls barrier \cite{Caillard2003}.  
Using the value $\Delta E_{\rm P}=25$\,meV/{\AA} given by this EAM potential 
for basal glide \cite{Chaari2014}, we obtain $\tau^*=440$\,MPa. 
This value is just a little bit lower than the stress, around 600\,MPa,
where a transition is observed in our simulations for the shape of the critical nucleus 
and for the variations of the activation enthalpy.
But, as it will be shown in the next subsection, the transition between 
the high- and the low-stress regime cannot be limited to a change of shape of the critical nucleus, 
as predicted by line tension models.
This transition also comes with an evolving atomic structure of the kinks,
with important consequences on the dislocation glide plane.

This critical nucleus defines the activation volume for basal glide, 
\ie{} how the activation enthalpy $\Delta H^{\rm act}$ depends on the applied stress. 
We calculate the area $\Delta A^{\rm act}_{xy}$ swept by this critical nucleus in the basal plane.
Results are shown in Fig. \ref{fig:activation_volume} with purple open squares for different applied stresses. 
This activation area should be equal to the first derivative of the activation enthalpy 
with the applied stress normalised by the norm of the Burgers vector
\begin{equation}
	\Delta A^{\rm act}_{xy}(\tau) = - \frac{1}{b} \frac{\partial \Delta H^{\rm act}}{\partial \tau}.
	\label{eq:actArea}
\end{equation}
The activation area defined by this equation is shown in Fig. \ref{fig:activation_volume} with red filled squares,
considering numerical derivation of the activation enthalpy given by NEB calculations (Fig. \ref{fig:activation_NEB}). 
A reasonable agreement is obtained between both definitions of activation areas, 
illustrating the consistency between the variations of the activation enthalpy
and the shape of the critical nucleus.
A peak of the activation area appears at 600\,MPa,
a signature of a change in the mechanism controlling dislocation mobility \cite{Caillard2003}.
The activation area can also be derived from the two Kocks laws 
used to interpolate the activation enthalpy, leading to two analytical laws
which are valid either in the low- or in the high-stress regime, 
but cannot reproduce the peak at 600\,MPa.

\subsection{Atomic structure of the kinks}

\begin{figure}[!bth]
	\begin{center}
		\includegraphics[width=0.99\linewidth]{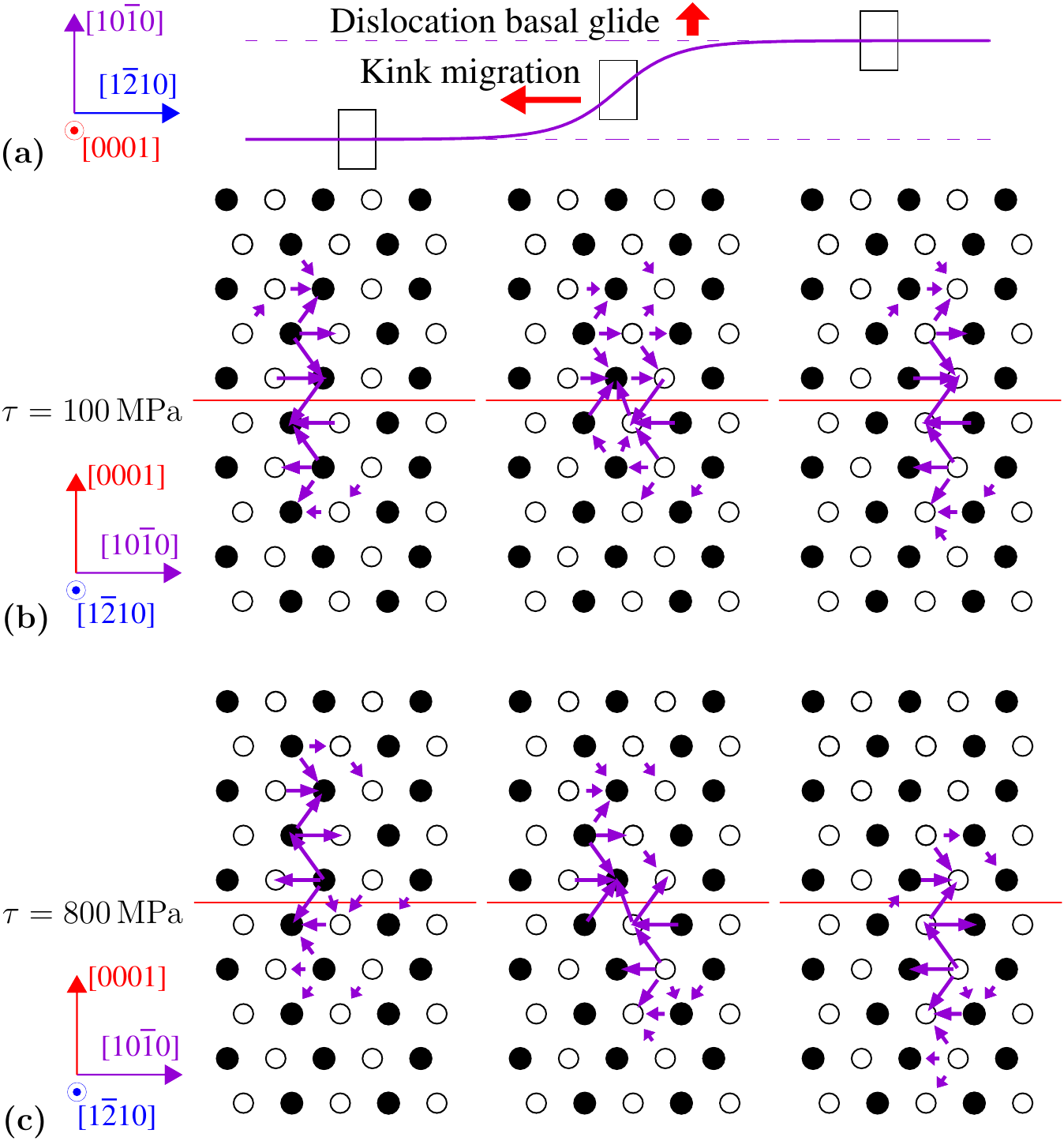}
	\end{center}
	\caption{Structure of the kinked dislocation for different applied stresses $\tau$.
	The continuous purple line in (a) represents the dislocations lying in two different Peierls valleys (purple dashed line)
	while gliding in a basal plane. 
	The differential displacement maps corresponding to the dislocation sections identified in (a) by black rectangles
	are shown in (b) for $\tau=100$\,MPa and in (c) for $\tau=800$\,MPa. 
	In these (b) and (c) figures, the left and right maps describe the structure of the dislocation 
	lying respectively in its initial and final Peierls valley, while the central map describes 
	the structure inside the kink.  The red line corresponds to the basal glide plane.}
	\label{fig:kink_struct}
\end{figure}

This change of the critical nucleus shape around 600\,MPa occurs with a modification 
of its atomic structure.  The configuration of the critical kink pair is shown 
in Fig. \ref{fig:kink_struct} for two different stresses, 100 and 800\,MPa,
using differential displacement maps to illustrate how the dislocation core varies 
along the dislocation line.
Far from the kink, the core adopts its ground state configuration and is dissociated in the prismatic plane. 
The parts on the left and on the right of the kink 
are lying in two different Peierls valleys separated by one minimal distance in the $\hkl[10-10]$ direction 
of basal glide.  At low stress (Fig. \ref{fig:kink_struct}b), these dislocation segments 
in the initial and arrival Peierls valleys have their centres located exactly in the same basal plane
which corresponds to the glide plane.  
In its kinked part, the dislocation adopts a configuration with a mixture of basal and prismatic spreading.
At low stress, the kinked dislocation is therefore fully lying in the basal plane where it is gliding.
The picture is different at higher stresses, as the dislocation segments in the initial and arrival Peierls valleys
have glided in their prismatic planes in opposite direction and are not located anymore in the same basal plane
(Fig. \ref{fig:kink_struct}c). The centres of these segments now belong to the same pyramidal plane. 
In agreement with these positions of the left and right parts of the dislocation, 
the core in the kinked part of the line is spread in this pyramidal plane.
NEB calculations therefore predict that the critical nucleus is lying in a pyramidal plane at high stress, 
thus explaining the pyramidal jumps observed in MD simulations when the dislocation is gliding in a basal plane.

One can legitimately wonder how this change of the kink structure with the applied stress 
depends on the choice of the interatomic potential. 
The EAM potential \#3 of Mendelev and Ackland \cite{Mendelev2007} is known to stabilize 
a metastable configuration of the straight screw dislocation 
corresponding to a dissociation in two Shockley partials in the basal plane,
whereas this basal structure is found unstable with \abinitio{} calculations \cite{Clouet2012}.
Although the kinked part of the screw dislocation partly spreads in the basal plane
in the low stress regime (Fig. \ref{fig:kink_struct}b), the atomic structure completely differs
from a full dissociation in two partials.  
It seems therefore unlikely that this structure of the kinked dislocation
is a consequence of the basal dissociation of the straight screw dislocation artificially stabilized 
by this empirical potential, especially as this metastable configuration has a much higher energy
than the ground state ($\Delta E=62$\,meV\,{\AA}$^{-1}$).
On the other hand, the kink structure at high stress (Fig. \ref{fig:kink_struct}c) looks similar to the dissociation
of the screw dislocation in two $1/6\,\hkl<1-210>$ partials in the pyramidal plane 
obtained not only with this empirical potential but also with \abinitio{} calculations
\cite{Chaari2014,Chaari2014a,Clouet2015}.

\subsection{Basal and pyramidal slip}

NEB calculations confirm that basal slip is composite at high stress
and can be decomposed in elementary prismatic and pyramidal slip events, 
in agreement with the glide mechanism already revealed
by \abinitio{} calculations for straight dislocations \cite{Chaari2014}.
But a change of the glide mechanism, corresponding to a change of the kink structure, is seen at low stress. 
At low stress, dislocation glide is fully resolved in the basal plane
and basal slip becomes an elementary slip system, thus in agreement 
with experimental observations \cite{Gong2015,Caillard2018,Wang2019}.

Basal and pyramidal slips appear thus intertwined.
To further demonstrate this link,
additional NEB calculations have been performed 
for a screw dislocation gliding in a pyramidal plane by nucleation and propagation
of a kink pair.  The setup is the same as for NEB calculations of basal glide
but with a different position of the dislocation final configuration
so as to lead to pyramidal slip
(see inset in Fig. \ref{fig:activation_NEB}
where the initial and final positions are shown with blue crosses).
Whatever the applied stress, NEB calculations lead to exactly the same activation enthalpy 
for pyramidal and basal slip (Fig. \ref{fig:activation_NEB}).
This can be rationalised by looking in more details to the mechanism for pyramidal slip. 
At low stress, the dislocation glide is composite and proceeds trough 
nucleation of a kink pair in the basal plane preceded and followed by easy slip 
of the screw dislocation in its initial and arrival prismatic plane.
On the other hand, at high stress the dislocation glide by nucleation of a kink pair
directly in the pyramidal plane.

Basal and pyramidal glide of $1/3\,\hkl<1-210>$ screw dislocations
are therefore two facets of the same  deformation mode,
with basal and pyramidal glide being favoured respectively at low and high stress. 
Experimental conditions usually correspond to the low stress regime, 
thus explaining why basal glide is the secondary slip system usually reported 
for $1/3\,\hkl<1-210>$ dislocations
\cite{Bailey1962,Dickson1971,Akhtar1973,Francillette1997,Francillette1998,Long2015a,Caillard2018}.
In particular, this preference for basal slip at low stress is fully compatible with the TEM observations
of Caillard \etal{} \cite{Caillard2018} during \insitu{} tensile experiments 
showing that basal glide is an elementary slip system and not a combination 
of two different slip systems.
Experimental evidences of pyramidal slip also exist. 
In particular, Tenckhoff \cite{Tenckhoff2005} mentioned that ``in regions of high stress concentrations,
such as grain boundaries, \hkl{10-11} slip traces occur'', thus in perfect agreement 
with the evolution of slip activity with the applied stress deduced from our NEB calculations.

\subsection{Comparison with experiments}

A good agreement is therefore obtained between atomistic simulations 
and \insitu{} TEM straining experiments \cite{Caillard2018}, 
both showing that basal glide is the secondary slip system  activated at low stress for \hkl<a> dislocations
and that it is controlled by a Peierls mechanism, \ie{} the nucleation
and propagation of kink pairs on screw dislocations.
This basal glide activated below 600\,MPa agrees
with compression experiments of micropillars realised by Wang \etal{} \cite{Wang2019}
on single-crystals with a high Schmid factors on basal slip systems, 
as these experiments show only basal slip traces, without any evidence of pyramidal slip, 
for resolved shear stresses ranging from 50 to 150\,MPa.
Besides this agreement on the activation of basal slip at low stress and the corresponding Peierls mechanism,
it is worth looking how the mobility law extracted from these simulations (Eq. \ref{eq:velocity_final}) 
and the obtained parameters compare with similar laws deduced from experiments \cite{Akhtar1973,Wang2019}.

Akhtar \cite{Akhtar1973} fitted a Norton law on tensile tests experiments on single crystals 
oriented to activate basal slip in a temperature range 850-1100\,K. 
The activation energy he obtained, 1.48\,eV, is lower than the one given by NEB calculations 
in the limit of a vanishing stress, 2.2\,eV (Fig. \ref{fig:activation_NEB}).
Such a difference is not surprising 
as the empirical potential used for the atomic simulations is known to overestimate 
the Peierls energy barrier for basal glide of a straight dislocation:
this potential leads to a 25\,meV/{\AA} barrier
whereas the barrier predicted by \abinitio{} calculations is only 9\,meV/{\AA} \cite{Chaari2014}.
One therefore expects that the activation energy controlling basal glide in zirconium
is lower than the value, 2.2\,eV, given by this potential.

Akhtar \cite{Akhtar1973} also measured activation volumes. 
The obtained values are above $200\,b^3$, much higher than the ones around $10\,b^3$
given by the atomic simulations.  As noticed by Akhtar such high values are not compatible
with a Peierls mechanism. 
He proposed that basal glide would operate by a cross-slip mechanism. 
Such a mechanism does not correspond to the one revealed by atomistic simulations. 
Besides, it would necessitate that the screw dislocation can adopt a metastable configuration
which can easily glide in basal planes.  
The best configuration would be a core dissociated
in a basal plane, but \abinitio{} calculations have shown that such a basal core is unstable \cite{Clouet2012}.
Finally this cross-slip mechanism would lead to jerky and intermittent motion of screw dislocations, 
in contradiction with the smooth motion seen in TEM during \insitu{} straining experiments \cite{Caillard2018}. 
Therefore, the high value obtained by Akhtar for the activation volume 
is more likely the signature of another mechanism limiting basal glide
in the high temperatures explored in these experiments.  
Impurities, which should be mobile at these temperatures, 
may impede dislocation motion, leading to a hardening contribution which would 
exceed the Peierls mechanism controlling basal glide in pure zirconium.

Both Caillard \etal{} \cite{Caillard2018} and Wang \etal{} \cite{Wang2019}
indeed mentioned that basal glide becomes athermal at 573\,K, 
thus well below the temperatures in Akhtar experiments where dislocations should not 
feel anymore the lattice friction, even when gliding in basal planes. 
An approximate athermal temperature can be estimated from the mobility law (Eq. \ref{eq:velocity_final})
extracted from atomic simulations (See Eq. 29 in Ref. \cite{Clouet2021})
using parameters obtained in the low stress regime for the Kocks law 
and assuming that parameters $T_{\rm m}$ and $\nu$ are the same 
as the ones given by MD in the high stress regime. 
For reasonable values of the dislocation density ($10^{10}$--$10^{16}$\,m$^{-2}$)
and of the strain rate ($10^{-3}$--$10^{-1}$\,s$^{-1}$), 
the obtained athermal temperature is between 450 and 570\,K, 
thus further confirming that the lattice friction opposing glide 
should have a negligible effect in the high-temperatures regime 
explored by Akhtar \cite{Akhtar1973}.

\section{Conclusion}

Atomistic simulations show that \hkl<a> screw dislocations in zirconium can glide 
in basal and pyramidal planes while remaining dissociated in the prismatic plane. 
The motion proceeds through the nucleation and propagation of kink-pairs
and is thus thermally activated.  
The activation enthalpy entering dislocation mobility law, 
and its temperature dependence, are well described by a Kocks law and the Meyer-Neldel compensation rule. 
These simulations confirm that basal and pyramidal slip are intimately intertwined
with two different stress regimes. 
At high stress, pyramidal glide of the screw dislocation
is the elementary event controlling these two secondary slip systems 
and basal glide is a combination of easy prismatic glide and pyramidal glide.
One thus recovers the mechanism  already revealed by \abinitio{} calculations 
for the glide of a straight screw dislocation \cite{Chaari2014}.
On the other hand, at low stress, kinks nucleate in the basal plane, 
allowing for a glide of the screw dislocation fully resolved in this basal plane,
and pyramidal slip becomes composite. 
Basal glide appears thus as an elementary slip system in hcp zirconium at low stress,
in agreement with experiments showing well defined slip traces in the basal planes
\cite{Gong2015,Caillard2018}.

With this new understanding of the mechanisms controlling basal and pyramidal slip 
in pure zirconium, it appears possible now to study the impact of the different 
alloying elements on the activity of these secondary slip systems. 
As described in the introduction, pyramidal slip of \hkl<a> dislocations 
appear more likely in zirconium alloys than in pure zirconium.  
Among the different addition elements met in zirconium alloy, 
oxygen appears as one with the most prominent effect, 
with oxygen promoting dislocation cross-slip in pyramidal planes \cite{Baldwin1968,Chaari2017}.

Despite the difference in the ground state of the \hkl<a> screw dislocation, 
spread on the prismatic plane for zirconium and on the first order pyramidal plane for titanium \cite{Clouet2015},
secondary slip in the basal and pyramidal planes is similar in both metals above room temperature. 
Basal slip appears also easier than pyramidal slip in titanium, 
with screw dislocations gliding through a slow and viscous motion compatible with a Peierls mechanism \cite{Caillard2018}.
As \abinitio{} calculations have shown that a \hkl<a> screw dislocation dissociated in a basal plane 
is unstable in titanium \cite{Kwasniak2019},
basal glide in titanium is also realized without any basal dissociation. 
The same mechanism, where the screw dislocation glides in a basal planes trough the nucleation and propagation of kink-pairs
while remaining dissociated in another plane, either prismatic or pyramidal, is therefore expected in titanium.

\vspace{0.5cm}
\linespread{1}
\small

\textbf{Acknowledgements} -
The authors thank Thomas Swinburne and Mihai-Cosmin Marinica 
for fruitful exchanges on entropy contributions to activation free enthalpy,
and Daniel Caillard for long-standing discussions on basal glide in zirconium 
and on Peierls mechanism.
This work was performed using HPC resources from GENCI-CINES and -TGCC (Grants 2020-096847)
and is funded by the project Transport \& Entreposage of the French Institut Tripartite CEA-EDF-Framatome.

\section*{References}
\bibliographystyle{elsarticle-num}
\bibliography{maras2021}

\end{document}